\documentclass[prd,nofootinbib,floats,superscriptaddress,eqsecnum,tightenlines,11pt]{revtex4}
\usepackage{hyperref}
\usepackage{graphicx}
\usepackage{amsmath,amssymb,amsfonts,amsthm,latexsym,stmaryrd}
\usepackage{marginnote}
\usepackage{color}
\usepackage{subfigure}

\def\beq{\begin{equation}}
\def\eeq{\end{equation}}
\newcommand{\bea}{\begin{eqnarray}}
\newcommand{\eea}{\end{eqnarray}}
\def\bi{\begin{itemize}}
\def\ei{\end{itemize}}
\def\ba{\begin{array}}
\def\ea{\end{array}}
\def\bfig{\begin{figure}}
\def\efig{\end{figure}}
\def\gam{{}^\gamma \!}

\begin{document}

\title{Gravity as an SU(1,1) gauge theory in four dimensions}
\author{Hongguang Liu}
\email{liu.hongguang@cpt.univ-mrs.fr}
\affiliation{Centre de Physique Th\'eorique (UMR CNRS 7332) , Universit\'es d'Aix-Marseille et de Toulon, 13288 Marseille, France}
\affiliation{Laboratoire de Math\'ematiques et Physique Th\'eorique (UMR CNRS 7350), Universit\'e Fran\c cois Rabelais, Parc de Grandmont, 37200 Tours, France}
\affiliation{Laboratoire Astroparticule et Cosmologie, Universit\'e Denis Diderot Paris 7, 75013 Paris, France}
\author{Karim Noui}
\email{karim.noui@lmpt.univ-tours.fr}
\affiliation{Laboratoire de Math\'ematiques et Physique Th\'eorique (UMR CNRS 7350), Universit\'e Fran\c cois Rabelais, Parc de Grandmont, 37200 Tours, France}
\affiliation{Laboratoire Astroparticule et Cosmologie, Universit\'e Denis Diderot Paris 7, 75013 Paris, France}

\date{\today}

\begin{abstract}
We start with the Hamiltonian formulation of the first order action of pure gravity with a
 full $\mathfrak{sl}(2,\mathbb C)$ internal gauge symmetry.
We make a partial gauge-fixing which reduces $\mathfrak{sl}(2,\mathbb C)$ to its sub-algebra $\mathfrak{su}(1,1)$. This case
corresponds to a splitting of
the space-time ${\cal M}=\Sigma \times \mathbb R$ where $\Sigma$ inherits an arbitrary Lorentzian metric
of signature $(-,+,+)$. Then, we find a parametrization of the phase space in terms of an $\mathfrak{su}(1,1)$ commutative
connection and its associated conjugate electric field. Following the techniques of Loop Quantum Gravity, we start the
quantization of the theory and we consider  the kinematical Hilbert space on a given fixed graph $\Gamma$ whose edges are colored
with unitary representations of $\mathfrak{su}(1,1)$. We compute the spectrum of area operators acting of the kinematical
Hilbert space: we show that
space-like areas have discrete spectra,  in agreement with usual $\mathfrak{su}(2)$ Loop Quantum Gravity,  whereas time-like areas have continuous spectra. We conclude on the possibility to make use of
this formulation of gravity to construct a holographic description of black holes in the framework of Loop Quantum Gravity.
\end{abstract}

	\maketitle

	\section{Introduction}
Loop quantum gravity was founded on the observation by Ashtekar \cite{ashtekar-variables} that working only with the self-dual part (or
equivalently the anti-self-dual part) of the Hilbert-Palatini action leads to a simplified parametrization of the phase space of pure gravity. Indeed,
the canonical variables are very similar to those of Yang-Mills gauge theory, there is no second class constraints and the first
class constraints associated to the local symmetries are polynomial functionals of the the canonical variables.
The drawback of the original Ashtekar's approach is that the phase space becomes complex  and then one requires the imposition of reality conditions in order to recover  the phase space of real general relativity. Of course, if one imposes the reality conditions at the classical level, prior to quantization, one looses all the beauty of the Ashtekar formulation, and recovers the standard Palatini formulation of general relativity, which we do not know how to quantize. Unfortunatelly, so far no one knows how to go the other way around, and implement the reality conditions after quantization of the Ashtekar theory.
This difficulty motivated the work of Barbero \cite{barbero} and, later on, Immirzi \cite{immirzi}, who introduced a family of canonical transformations, parametrized by the so-called Barbero-Immirzi parameter $\gamma$, and leading to a canonical theory in terms of a real $\mathfrak{su}(2)$ connection kown as the Ashtekar-Barbero connection. The action that leads to this canonical formulation was finally found by Holst \cite{holst}.

In fact, the Holst action is a first order formulation of gravity with a full $\mathfrak{sl}(2,\mathbb C)$ internal symmetry and an explicit dependency
on the parameter $\gamma$ which appears as a coupling constant for a topological term. One uses a partial gauge fixing in this action
in order to derive a canonical theory in terms of the Ashtekar-Barbero.  This choice of gauge  is referred to as the time gauge, and, by doing so, the Lorentz gauge algebra in the internal space is reduced to its rotational $\mathfrak{su}(2)$ subalgebra. Finally, Loop Quantum Gravity is a
canonical quantization of  this gauge fixed first order formulation of gravity which lead to a beautiful construction of the space of
quantum geometry states at the kinematical level. At this stage, one can naturally ask the question whether
the construction of Loop Quantum Gravity deeply relies on the time gauge or not. A related question would be whether the physical
predictions of Loop Quantum Gravity are changed or not when one makes another partial gauge fixing or no gauge fixing at all
in the Holst action prior to quantization. Indeed, the discreteness of the quantum geometry at the Planck scale predicted in Loop Quantum
Gravity can be interpreted  as a direct consequence of the compactness (via Harmonic analysis) of the residual symmetry group $SU(2)$
in the time gauge. These important problems have been studied quite a lot the last twenty years but it is fair to say that no definitive
conclusion have closed the debates so far.

Most of the approaches to address this issue are based on attempts to quantize the
Holst action without any partial gauge fixing, and then keeping the full Lorentz internal invariance of the theory.
Now if one performs the canonical analysis of the $\mathfrak{sl}(2,\mathbb C)$ Holst action, second class constraints appear simply because the connection has more components than the tetrad field. The appearance of second class constraints makes the classical analysis and then the quantization of the theory much more involved.
In the analysis of constrained systems, there are two ways of dealing with second class constraints: one can either solve them explicitly, or implement them in the symplectic structure by working with the Dirac bracket.
These two methods are totally equivalent.
Using the Dirac bracket, Alexandrov and collaborators \cite{alexandrov1,alexandrov2,alexandrov3,alexandrov4}
were able to construct a two-parameters family
of Lorentz-covariant connections (which are diagonal under the action of the area operator, and transform properly under the action of spatial diffeomorphisms). Generically, these connections are non-commutative and therefore the theory becomes very difficult to quantize.
The alternative route to deal with covariant connections
was initiated by Barros e Sa in \cite{barros} who solved explicitly the second class constraints.
In this approach, the phase space is parametrized by two pairs of canonical variables:
the generalization $(A,E)$ of the usual Ashtekar-Barbero connection and its conjugate densitized triad $E$;
and a new pair of canonically conjugated fields $(\chi,\zeta)$, where $\chi$ and $\zeta$ both take values in $\mathbb R^3$. Then,
Barros e Sa expressed the remaining boost, rotation, diffeomorphism and scalar constraints in terms of these variables. The elegance of this approach is that it enables one to have a simple symplectic structure with commutative variables, and a tractable expression for the boost, rotation and diffeomorphism generators. Although the scalar constraint becomes more complicated, this structure is enough to study the
kinematical structure of loop quantum gravity with a fully Lorentz invariance. This has  precisely been done in \cite{geiller1,geiller2} where
one constructed the unique spatial connection which is not only commutative but also
transforms covariantly under the action of  boosts and rotations. In fact, this connection coincides with the
commutative Lorentz connection studied earlier in \cite{alexandrov4} and
the one found in \cite{les italiens}. Furthermore, it has been shown to be gauge related to the Ashtekar-Barbero connection
via a pure boost parametrized by the vector $\chi$ viewed as a velocity. Hence, the construction proposed in  \cite{geiller1,geiller2}
works only when $\chi^2 < 1$. Thus, the pairs of canonical variables formed with the $\mathfrak{sl}(2,\mathbb C)$ connection and its conjugate electric field parametrize only a part of the fully covariant phase space of the Holst action.

 This paper enables us to explore
the sector $\chi^2 >1$ while studying a partial gauge fixing of the Holst action that reduces $\mathfrak{sl}(2,\mathbb C)$ to $\mathfrak{su}(1,1)$.
Hence, we start with the Lorentz covariant parametrization of the Holst action found by Barros e Sa \cite{barros}.  We find
a partial gauge fixing which breaks the $\mathfrak{sl}(2,\mathbb C)$ internal symmetry into $\mathfrak{su}(1,1)$ and this  is  possible
if and only if $\chi^2 >1$.  Such a partial gauge fixing corresponds to a canonical
splitting of the space-time  ${\cal M}=\Sigma \times \mathbb R$ where $\Sigma$ is no more space-like
(as it is the case in the usual Ashtekar-Barbero parametrization) but inherits a Lorentzian metric  of signature is $(-,+,+)$.
As a consequence, only three out of the initial
six first class constraints remain after the partial gauge fixing, and they generate as expected the local $\mathfrak{su}(1,1)$
gauge transformations. The other three constraints  form with the three gauge fixing conditions a set of second class constraints
that we solve explicitly.
Then, we construct an $\mathfrak{su}(1,1)$ connection which appears to be commutative in the sense of the
Poisson bracket. This remarkable construction allows us to investigate the loop quantization of the theory  and to build the kinematical
Hilbert space on a given graph $\Gamma$ whose edges are associated to $SU(1,1)$ holonomies.  It is well-known that \cite{freidel livine}
the non-compactness of the gauge group prevents us from defining the projective limit of spin-networks and then the sum over all graphs
of kinematical Hilbert space is ill-defined. Nonetheless, if one restricts the study to one given graph $\Gamma$,
it is possible to define the action of the
area operator and one easily finds that a space-like area has a discrete spectrum whereas the spectrum of a time-like area is continuous.
In other words, if one considers a spin-network defined on a graph $\Gamma$ dual to a discretization $\Delta=\Gamma^*$
of a $(2+1)$-dimensional manifold, edges $e$ of $\Gamma$ are colored with representations in the discrete series (resp. in the continuous
series) if the dual face $f=e^*$ of $\Delta$ is space-like (resp. time-like).  The spectrum of space-like areas is in total agreement with the one  obtained in the usual Ashtekar-Barbero formalism for space-like surfaces.

The paper is organized as follows. After the introduction, we start in Section II with a brief summary of the canonical analysis \`a la Barros e sa
of the fully Lorentz invariant Holst action. In Section III, we present the partial gauge fixing that breaks $\mathfrak{sl}(2,\mathbb C)$ into
$\mathfrak{su}(1,1)$ before constructing the $\mathfrak{su}(1,1)$ connection and its associated electric field. In Section IV we explore  the kinematical
quantization of the theory on a given graph and we compute the spectra of area operators which act unitarily in the kinematical Hilbert space.
We conclude in Section V with a brief summary of the most important results and a discussion on the consequences of this new parametrization for the description of black holes in Loop Quantum Gravity.

	\section{First order Lorentz-covariant gravity}

In this section, we summarize the main results of the Hamiltonian analysis of the fully Lorentz invariant Holst action.
We start recalling  the main steps of the constraints analysis and present the solutions of the second
class constraints proposed by Barros e Sa  \cite{barros}. Then, we describe
the parametrization of the Lorentz covariant phase space that will serve  to build the $\mathfrak{su}(1,1)$ connection in the next
Section. Finally,  we discuss the structure of  the first class constraints focussing mainly on the generators of the internal
Lorentz symmetry.

\subsection{Action and constraints analysis}

The Holst action \cite{holst} is a generalization of the Hilbert-Palatini first order action with a Barbero-Immirzi parameter $\gamma$.
In terms of the co-tetrad $e^I_\alpha(x)$ and the Lorentz connection one-form $\omega^{IJ}_\alpha(x)$, the
corresponding Lagrangian density is
\bea\nonumber
\mathcal{L}[e,\omega]=  \,\frac{1}{2} \epsilon_{IJKL}  \, e^I\wedge e^J\wedge F^{KL} \, + \,
\frac{1}{\gamma} e^I\wedge e^J\wedge F_{IJ} \, ,
\eea
where $F[\omega]=d\omega+\omega\wedge\omega$ is the curvature two-form of the connection $\omega$, $\epsilon_{IJKL}$
the fully antisymmetric symbol which defines an invariant non-degenerate bilinear form on $\mathfrak{sl}(2,\mathbb C)$, and internal
indices are lowered and raised with the flat metric $\eta_{IJ}$ and its inverse $\eta^{IJ}$.
It is well known that the Holst action is equivalent to the Hilbert-Palatini action. Indeed, if the co-tetrad is not degenerated (i.e. if its determinant is not vanishing), one can uniquely solve $\omega$ in terms of $e$ (from the torsionless equation for $\omega$)
and find that $\omega_\mu^{IJ}$ are nothing but the components of
the Levi-Civita connection. Plugging back this solution into the action eliminates the Barbero-Immirzi parameter $\gamma$
by virtue of the Bianchi identities and leads to the second order Einstein-Hilbert action.

\medskip

Now, we recall basic results on the canonical analysis of the Holst Lagrangian. For this purpose, it is convenient to introduce the notation
\bea\label{gam}
{}^\gamma \! \xi_{IJ}=  \xi_{IJ }- \frac{1}{2\gamma} \epsilon_{IJKL} \, \xi^{KL}\, ,
\eea
for any element $\xi \in \mathfrak{sl}(2,\mathbb C)$. After performing a $3+1$ decomposition (based on a splitting
${\cal M}={\Sigma} \times \mathbb R$ of the space-time) in order to distinguish between temporal and spatial
coordinates ($0$ is the time label and small latin letters from the beginning of the alphabet $a,b,c,\cdots$ hold for spacial indices),
a straightforward calculation leads to the following canonical expression of the Lagrangian density
\bea
\mathcal{L}[e,\omega]=\gam\pi^a_{IJ}\, \dot{\omega}^{IJ}_a-g^{IJ}\mathcal{G}_{IJ}-N\mathcal{H}-N^a\mathcal{H}_a,\label{lagrangian}
\eea
where we have introduced the notations  $\dot{\omega}=\partial_0\omega$  for the time derivative of $\omega$, $g^{IJ}$ for $-\omega^{IJ}_0$, $N$ for the lapse function $N$, and $N^a$ for the shift vector.
All these functions are Lagrange multipliers which enforce respectively the Gauss, Hamiltonian, and diffeomorphism constraints
\bea\label{constraints}
\mathcal{G}_{IJ}=D_a\gam\pi^a_{IJ},\qquad\qquad
\mathcal{H}=\pi^a_{IK}\pi^{bK}_J\gam F^{IJ}_{ab},\qquad\qquad
\mathcal{H}_a=\pi^b_{IJ}\gam F^{IJ}_{ab}.
\eea
These constraints are expressed in terms of the spatial connection components $\omega_a^{IJ}$, and the canonical momenta defined by
\bea\label{pi=ee}
\pi^a_{IJ}\equiv \epsilon_{IJKL}\, \epsilon^{abc}\, e^K_be^L_c.
\eea
Since $\pi^a_{IJ}=-\pi^a_{JI}$ contains 18 components, and the co-tetrad has only 12 independent components, we need to impose 6 primary constraints often called the simplicity constraints
\bea\label{simplicity}
\mathcal{C}^{ab}=\epsilon^{IJKL}\pi^a_{IJ}\pi^b_{KL}\approx0,
\eea
in order to parametrize the space of momenta in terms of the $\pi$ variables instead of the co-tetrad variables.
Classically, it is equivalent to work with the 12 components $e_a^I$ or with the 18 components $\pi^a_{IJ}$ constrained to satisfy the 6 relations $\mathcal{C}^{ab}\approx0$. Hence, at this stage, the non-physical Hamiltonian phase space is  parametrized by the 18 pairs of canonically conjugated variables $(\omega_a^{IJ},\pi^a_{IJ})$, with the set of 10 constraints (\ref{constraints}) to which we add  the 6
constraints $\mathcal{C}^{ab}\approx0$.

Studying the stability under time evolution of these ``primary" constraints is rather standard
and has been performed first for the Hilbert-Palatini action in \cite{peldan} and for the Holst action in \cite{barros}.
Here we will not reproduce all the steps of this analysis, but only focus on the structure of the second class constraints and their resolution.
Details with our notations can be found in \cite{geiller1}.
Notice first that in order to recover the $4$ phase space degrees of freedom (per space-time points) of gravity, the theory needs to have secondary constraints, which in addition have to be second class. This is indeed the case. Technically, this comes
from the fact that the algebra of constraints fails to close because the scalar constraint $\mathcal{H}$ does not commute weakly with the simplicity constraint $\mathcal{C}^{ab}$. Hence, requiring their stability under time evolution  generates the following
6 additional secondary constraints
\bea\nonumber
\mathcal{D}^{ab}=\epsilon_{IJ MN}\, \pi^{cMN}\left(\pi^{aIK}D_c\pi^{bJ}_{~~K}+\pi^{bIK}D_c\pi^{aJ}_{~~K}\right)\approx0.
\eea
The Dirac algorithm closes here with $18\times 2$ phase space variables (parametrized by the components of $\pi$ and $\omega$), and
22 constraints $\mathcal{H}$, $\mathcal{H}_a$, $\mathcal{G}_{IJ}$, $\mathcal{C}^{ab}$ and $\mathcal{D}^{ab}$. Among these constraints,
the first 10 are first class (up to adding second class constraints) as expected, and the remaining 12 are second class. One can check explicitly that
$\mathcal{C}^{ab}\approx0$ and $\mathcal{D}^{ab}\approx0$ form a set of second class constraints (their associated Dirac matrix is invertible), and that the first class constraints generate the symmetries of the theory, namely the space-time diffeomorphisms and the Lorentz gauge symmetry. Finally,
we are left with the expected 4 phase space degrees of freedom per spatial point:
\bea
18 \times 2 \text{(dynamical variables)} - 10 \text{(first class constraints)}\times 2 - 12 \text{(second class constraints)}. \nonumber
\eea
We recover the two gravitational modes.

\subsection{Parametrization of the phase space}

Now that we have clarified the Hamiltonian structure of the theory, we are going to show how to solve  the second class constraints
following  \cite{barros}. First, one writes the 18 components of $\pi_a^{IJ}$ as
\bea\label{solution}
\pi^a_{0i}={2}E^a_i,\qquad\qquad\pi^a_{ij}={2}(E^a_{i}\chi_{j} - E^a_{j}\chi_{i} )\, ,
\eea
where $\chi_i = e^a_i  e_a^0$ (which encodes the deviation of the normal to the hypersurfaces from the time direction) and $E^a_i$
(which corresponds to the usual densitized triad of loop gravity) are now twelve independent variables. Note that $e^a_i$ is the
inverse of $e_a^i$ viewed as a $3\times 3$ matrix. This is trivially a solution
of the simplicity constraints \eqref{simplicity} because somehow we have returned to the co-tetrad parametrization \eqref{pi=ee}.

Then, we plug the solution (\ref{solution}) into the canonical term of the Lagrangian (\ref{lagrangian}) which gives
\bea\label{new variables}
\gam\pi^a_{IJ}\dot{\omega}^{IJ}_a=E^a_i\dot{A}^i_a+\zeta^i\dot{\chi}_i \qquad
\text{where} \quad
A^i_a=\gam\omega^{0i}_a+\gam\omega^{ij}_a\chi_j\quad \text{and} \qquad\zeta^i=\gam\omega^{ij}_aE^a_j.
\eea
This result strongly suggests  that the 18 components of the connection could be expressed in terms of the  12
independent variables $(A_a^i,\zeta^i)$ when one solves the 6 secondary second class constraints.  This is indeed the case and it
 can be seen by inverting the relation (\ref{new variables}) as follows
\bea\label{invertion}
\gam\omega^{0i}_a=A^i_a-\gam\omega^{ij}_a\chi_j,\qquad\qquad\gam\omega^{ij}_a=\frac{1}{2}\left(Q^{ij}_a-E_a^{i}\zeta^{j}
-E_a^{j}\zeta^{i}  \right),
\eea
where $E^i_a$ is the inverse of $E^a_i$, and $Q_a^{ij}=Q_a^{ji}$ has a vanishing action on $E^a_i$. The explicit form of $Q_a^{ij}$
can be obtained from ${\cal D}^{ab} \approx 0$ as shown in \cite{barros}. Furthermore, when $\gamma^2 \neq 1$, one can uniquely
express $\omega$ in terms of $\gam\omega$ using the inverse of the map \eqref{gam}.

As a consequence,
the phase space can be parametrized by the twelve pairs of canonical variables $(A^i_a,E^a_i)$ and $(\chi_i,\zeta^i)$ with the (non-trivial)  Poisson brackets  given by
\bea\label{phase space}
\big\{A^i_a(x),E^b_j(y)\big\}=\delta^i_j\delta^b_a\, \delta^3(x-y)\quad \text{and} \quad
\big\{\chi_i(x),\zeta^j(y)\big\}=\delta^j_i \, \delta^3(x-y).
\eea
Remark that if we work in the time gauge (i.e. $\chi=0$), the variable $A^i_a$ coincides exactly with the usual Ashtekar-Barbero connection.

\subsection{First class constraints}

It remains  to express the first class constraints (\ref{constraints}) in terms of the new phase space variables (\ref{phase space}). This is
an easy task using  the defining relations \eqref{solution} and \eqref{invertion}. This  was done by  Barros e Sa.
The constraints have quite a simple form except the Hamiltonian
constraint whose expression is more involved: it can be found in \cite{barros} and we will not consider this constraint in this paper.
The vector constraint $\mathcal{H}_a$ takes the form
\bea
\mathcal{H}_a &=& E^b\cdot (\partial_{a}A_{b} - \partial_b A_a) +\zeta\cdot\partial_a\chi+\frac{\gamma^2}{1+\gamma^2}\Big[(E^b\cdot A_b)(A_a\cdot\chi)-(E^b\cdot A_a)(A_b\cdot\chi)\nonumber\\
&&+(A_a\cdot\chi)(\zeta\cdot\chi)-(A_a\cdot\zeta)+\frac{1}{\gamma}\left(E^b\cdot(A_b\times  A_a)+\zeta\cdot(\chi \times A_a)\right)\Big] \, ,\label{vector constraint}
\eea
where $\cdot $ denotes the scalar product  $\lambda \cdot \mu=\lambda_i \mu^i$ and $\times$ denotes
the cross product $(\lambda \times \mu)^i=\epsilon^{ijk}\lambda_j \mu_k$ for any two pairs of vectors $\lambda$ and $\mu$ in $\mathbb R^3$.
Concerning, the Lorentz constraints $\mathcal{G}_{IJ}$, they can be split into its boost part $\mathcal{B}_i\equiv\mathcal{G}_{0i}$, and its rotational part $\displaystyle\mathcal{R}_i\equiv(1/2)\epsilon_i^{~jk}\mathcal{G}_{jk}$ whose expressions are
\begin{subequations}\label{gauss}
\bea
\mathcal{B}&=&\partial_a\left(E^a-\frac{1}{\gamma}\chi\times E^a\right)- (\chi\times E^a)\wedge A_a+\zeta-(\zeta\cdot\chi)\chi \,,\\
\mathcal{R}&=&-\partial_a \left(\chi\times E^a+\frac{1}{\gamma}E^a\right)+A_a\times E^a-\zeta\times\chi \, .
\eea
\end{subequations}
One can check  that these constraints  satisfy indeed the Lorentz algebra
\bea
\{\mathcal{B}\cdot u,\mathcal{B} \cdot v \}=-\mathcal{R} \cdot u\times v,\quad
\{\mathcal{R}\cdot u,\mathcal{R}\cdot v\}=\mathcal{R}  \cdot u\times v,\quad
\{\mathcal{B}\cdot u,\mathcal{R}\cdot v\}=\mathcal{B}  \cdot u\times v \, ,
\eea
where $u$ and $v$ are arbitrary vectors.

In the time gauge, one immediately recovers the constraints structure of the formulation of gravity in terms of the
Astekar-Barbero connection. In that case, $\chi \approx 0$ drastically simplifies the boost constraints which become
equivalent to $\zeta-\partial_aE^a \approx 0$. The conditions  $\chi \approx 0$ and $\zeta-\partial_aE^a \approx 0$
form a set of second class constraints that can be solved explicitly for $\chi$
and $\zeta$. By doing so, the variables $(\chi,\zeta)$ are eliminated from the theory, and the vectorial, the rotational and also the
Hamiltonian constraints are those of Loop Quantum Gravity.

Now our task is to use the phase space variables (\ref{phase space}) to make a partial gauge fixing which reduces
the original Lorentz algebra to $\mathfrak{su}(1,1)$.

	\section{Gravity as an SU(1,1) gauge theory}

In this section, we first show how to make a partial gauge fixing of the full Lorentz invariant Holst action which reduces the internal
$\mathfrak{sl}(2,\mathbb C)$ gauge symmetry to $\mathfrak{su}(1,1)$. At the same time, we keep the invariance under
diffeomorphisms on $\Sigma$. In that case, we will see that the splitting of the space-time ${\cal M}= \Sigma \times \mathbb R$ is such that
$\Sigma$ is no more a space-like hypersurface as it is the case in the time gauge but inherits instead a Lorentzian structure.
Then, we construct a parametrization of the phase space in terms of an $\mathfrak{su}(1,1)$ connection and its conjugate
electric field which transforms in the adjoint representation of $\mathfrak{su}(1,1)$. Furthermore, we show that
these variables are Darboux coordinates for the phase space, which paves the way towards a quantization of the theory
explored in the following Section.

	\subsection{Breaking the internal symmetry: from $\mathfrak{sl}(2,\mathbb C)$ to $\mathfrak{su}(1,1)$}
	As we have already underlined in the previous section, imposing the time gauge $\chi \approx 0$ in the fully covariant Holst action
breaks the boost invariance and only the rotational parts of the constraints remain first class among the original 6 internal symmetries.
Hence, we get an $\mathfrak{su}(2)$ invariant theory of gravity. In fact, we proceed in a very similar way to construct an
$\mathfrak{su}(1,1)$ invariant theory from the Holst action: we find a partial gauge fixing such that two components of the boosts constraints
and one of the rotational constraints remain first class whereas the three others  form with the gauge fixing conditions a second class system.
Naturally, we consider a gauge fixing condition of the form
\bea\label{gauge fixing}
{\cal X} \equiv  \chi - \chi_0 \approx 0
\eea
where $\chi_0$ is a fixed non-dynamical vector. Inspiring ourselves with what happens in the time gauge, we expect \eqref{gauge fixing} to form a second  class system with three out of the six constraints \eqref{gauss}. These three second class components of the Lorentz generators are supposed to be
\bea\label{BRR}
{\cal R} \cdot u \approx 0 \, , \quad
{\cal R} \cdot v \approx 0 \, , \quad
{\cal B} \cdot n \approx 0 \, ,
\eea
where ${u}$ and ${v}$ are two given normalized orthogonal vectors and $n=v\times u$.
The reason is that we are left with two boosts and one rotations
which are expected to reproduce (up to the addition of second class constraints) an $\mathfrak{su}(1,1)$ Poisson algebra. To
derive the conditions for this  to happen, we start rewriting \eqref{BRR} as a linear system of equations for $\zeta$:
\bea\label{Mzeta}
M\zeta
=
\begin{pmatrix}
\zeta \cdot U  \\
\zeta \cdot V \\
\zeta \cdot W \\
\end{pmatrix}
\approx
\begin{pmatrix}
{\cal R} \cdot u\vert_{\zeta =0}  \\
{\cal R} \cdot v\vert_{\zeta =0} \\
 {\cal B} \cdot n\vert_{\zeta =0} \,
\end{pmatrix}
\quad \text{with}\quad
M \equiv
\begin{pmatrix}
{}^t U \\
{}^tV \\
{}^t W
\end{pmatrix}
\, \text{and} \,
\left\{
\begin{array}{lll}
U & \equiv &  \chi \times u \\
V & \equiv & \chi \times v \\
W & \equiv & -n + (\chi \cdot n) \chi
\end{array}
\right.
\eea
The system admits an unique solution for $\zeta$ if and only if
\bea\label{first condition}
\text{det} M = U \times V \cdot W = (1-\chi^2)  (\chi \cdot n)^2 \, \neq 0 \ ,
\eea
which implies that $\chi^2 \neq 1$ and $\chi \cdot n \neq 0$.
When we assume this is the case, the solution $\zeta_0$  can be easily expressed in terms of the components of
$\chi_0$, $E$ and $A$ inverting \eqref{Mzeta} as follows
\bea
\zeta_0 & = & M^{-1}
\begin{pmatrix}
{\cal R} \cdot u\vert_{\zeta =0}  \\
{\cal R} \cdot v\vert_{\zeta =0} \\
 {\cal B} \cdot n\vert_{\zeta =0} \,
\end{pmatrix}
 =
\frac{\left( {\cal B} \cdot n + {\cal R} \cdot \chi \times n \right) \chi - (1-\chi^2) {\cal R} \times n}{(1-\chi^2) n \cdot \chi} \vert_{\zeta=0}\, ,
\eea
where we used the expression
\bea
M^{-1} \; = \;
\frac{1}{U \times V \cdot W}
\begin{pmatrix}
V \times W  \, , \,  W \times U \, , \,  U \times V
\end{pmatrix}.
\eea
Hence, the three constraints \eqref{BRR} are equivalent to the three conditions
\bea\label{calZ}
{\cal Z} \equiv \zeta - \zeta_0(\chi_0,E,A) \approx 0 \, .
\eea
Now, it becomes clear that the gauge fixing conditions ${\cal X}\approx 0$ \eqref{gauge fixing} and the three constraints ${\cal Z} \approx 0$
form a second class system because their associated $6\times 6$ Dirac matrix $\Delta$
\bea
\Delta(x,y) \equiv
\begin{pmatrix}
X(x,y) & Y(x,y) \\
-{}^t Y(x,y) & Z(x,y)
\end{pmatrix} \, \text{with} \, \left\{
\begin{array}{lll}
X^i_j(x,y) & \equiv & \{ \chi^i(x), \chi_j(y) \} = 0 \\
Y^i_j(x,y) & \equiv & \{ {\cal X}^i(x),{\cal Z}_j(y)\} \, = \; \delta^j_i\delta^3(x-y) \\
Z^i_j(x,y) & \equiv & \{ {\cal Z}^i(x),{\cal Z}_j(y)\}
\end{array}
\right.
\eea
is invertible whatever $Z$ is. These two constraints allow to eliminate the variables $\chi$ and $\zeta$ from the phase space provided that one introduces the external non
dynamical field $\chi_0$.

\medskip

We are left with three constraints from \eqref{gauss} which are required to satisfy an $\mathfrak{su}(1,1)$ Poisson algebra
once one replaces $\chi$ by $\chi_0$ and $\zeta$ by $\zeta_0$. These constraints are denoted
 \bea\label{Jconstraints}
 {\cal J}_u \equiv {\cal B} \cdot u \vert_{\chi_0,\zeta_0} \, , \quad
 {\cal J}_v \equiv {\cal B} \cdot v \vert_{\chi_0,\zeta_0} \, , \quad
 {\cal J}_n \equiv {\cal R} \cdot n \vert_{\chi_0,\zeta_0}\, .
 \eea
 From now on, we will omit to mention the index $0$ for $\chi$ to lighten the notations. However,  $\chi$ has to be understood
 as an external non dynamical field, and not as the initial dynamical variable in the fully Lorentz invariant Holst action.

 A long but standard calculation shows that the three constraints \eqref{Jconstraints} form a closed Poisson algebra only when
\bea\label{second condition}
u \cdot \chi = v \cdot \chi = 0 \, .
\eea
This is equivalent to the condition that $\chi = \pm \vert \chi \vert n$ where $\vert \chi \vert \equiv \sqrt{\chi \cdot \chi}$ is the norm of
$\chi$. Without loss of generality, we choose  $\chi =\vert \chi \vert n$. As a consequence, the partial gauge fixing \eqref{gauge fixing} leaves the remaining three constraints \eqref{Jconstraints}
first class only when \eqref{second condition} is satisfied.  In that case, the expressions of \eqref{Jconstraints}
simplify a lot and   they can be written as
	\bea\label{projectJ}
	{\cal J}_0 \, \equiv \, {\cal{J}}_n  = n \cdot \tilde{\cal J} \, , \quad
	{\cal J}_1 \, \equiv \, C {\cal{J}}_v =  Cu  \cdot \tilde{\cal J} , \quad
	{\cal J}_2 \, \equiv \, C {\cal{J}}_u = \,  -Cv \cdot \tilde{\cal J} \, ,
	\eea
where $C=1/\sqrt{\vert \chi^2-1\vert}$ is a normalization function and we introduced the vector field
\bea
\tilde{\cal J} \; \equiv \;  -\frac{1}{\gamma}  \big( \partial_a E^a  +\partial_a (E^a \times \chi) \times \chi \big)+ \tilde{A}_a \times E^a
\eea
given in terms of  the $\mathfrak{su}(2)$-valued one form $\tilde{A}$  defined by
	\bea\label{tildeA}
	\tilde{A}_a=A_a-(A_a \cdot \chi) \chi -\partial_a \chi \, .
	\eea
Finally, one shows that the constraints algebra reduces to the simple form
	\bea
	\{ {\cal{J}}_0, {\cal{J}}_1 \}\,=\, {\cal{J}}_2 \, , \qquad
	\{ {\cal{J}}_0, {\cal{J}}_2 \}\,=\, -{\cal{J}}_1 \, , \qquad
	\{{\cal{J}}_1, {\cal{J}}_2 \} \,=\, \sigma {\cal{J}}_0 \, ,
	\eea
where
	\bea\label{sigma}
	 \sigma \equiv \frac{1-\chi^2}{\vert 1 - \chi^2 \vert} = \text{sg}(1-\chi^2).
	\eea
The function $\text{sg}(x)$ denotes the sign of $x \neq 0$.
As a consequence, the remaining three constraints form an $\mathfrak{su}(2)$ Poisson algebra when $\chi^2 <1$
and an $\mathfrak{su}(1,1)$ Poisson algebra when $\chi^2 >1$ (the case $\chi^2=1$ is excluded from the scope of our method and
should be studied in a different way\footnote{This case corresponds to a slicing of the space-time in a light like direction. Our analysis based
on a partial gauge fixing can be adapted to that situation. Such a Hamiltonian description could provide us with a new formulation
(eventually simpler) of gravity
in the light front related to \cite{simone}.}).
We can write the constraints algebra in the more compact form
\bea
\{ {\cal J}_\alpha , {\cal J}_\beta\} \; = \; \epsilon_{\alpha\beta}{}^\tau \, {\cal J}_\tau
\eea
where $\alpha,\beta,\tau \in (0,1,2)$ and $\epsilon_{\alpha\beta\tau}$ is the totally antisymmetric symbol with
$\epsilon_{012}=+1$. Furthermore, the indices are lowered and raised
with the flat metric and its inverse $\text{diag}(\sigma,+1,+1)$: it is the flat
 Euclidean metric $\delta_{\alpha\beta}$ when $\sigma = +1$ and  the flat Minkowski metric $\eta_{\alpha\beta} \equiv \text{diag}(-1,+1,+1)$
when $\sigma=-1$. Hence, as announced above, one recognizes respectively the $\mathfrak{su}(2)$ and the $\mathfrak{su}(1,1)$
Lie algebras.

\medskip

Let us close this analysis with one remark.
The gauge fixing condition \eqref{gauge fixing} makes the three constraints \eqref{BRR} (which are first class in the full Lorentz
invariant Holst action) second class. Hence, we have left two boosts and one rotation first class in order to get an $\frak{su}(1,1)$
gauge symmetry at the end of the process. This is what we arrive at when $\chi^2 >1$ but we obtain an $\mathfrak{su}(2)$ gauge
symmetry when $\chi^2 <1$ even though we kept two boosts among the remaining first class constraints.
The reason is that, at the end of the gauge fixing process, the remaining first class constraints are
non-trivial linear combinations of the six initial first class constraints and the gauge fixing conditions. Hence, they could form either
an $\mathfrak{su}(1,1)$ or an $\mathfrak{su}(2)$ algebra. The two most important ingredients in our construction is that, first,
we replace three out of the initial six first class constraints by constraints of the type \eqref{calZ} which fix $\zeta$,
and second we impose that the
remaining constraints (when $\zeta$ and $\chi$ are replaced from ${\cal X}\approx 0$ and ${\cal Z} \approx 0$)
form a closed Poisson algebra. In that respect, we could have considered the conditions
${\cal B}.u \approx {\cal B}.v \approx {\cal B}.n \approx 0$ instead of \eqref{BRR}: we would have obtained another set of conditions
fixing $\zeta$ and then, following the same strategy, we would have shown that the remaining three
constraints are generators of a closed algebra provided that \eqref{second condition} is satisfied. The remaining symmetry would have been
$\mathfrak{su}(2)$ or $\mathfrak{su}(1,1)$ depending on the sign of $\sigma$ exactly as in the previous analysis.

\subsection{On the space-time foliation}
Let us discuss the reason why the sign $\sigma$ of $(\chi^2-1)$ determines the signature of the symmetry algebra
$\mathfrak{su}(2)$ or $\mathfrak{su}(1,1)$.  For that purpose,
it is very instructive to study the properties  of the metric $g_{ab}$ induced on the hypersurface $\Sigma$ whose expression is
\bea\label{gab}
g_{ab} \equiv  e_a^I \eta_{IJ} e_b^J = e_a^i \gamma_{ij} e_b^j \quad \text{with} \quad  \gamma_{ij}  \equiv \delta_{ij} - \chi_i \chi_j
\eea
where we inverted the defining relation $\chi_i = e^a_i e_a^0$ to replace $e_a^0$ by $e_a^i \chi_i$. It is immediate to notice that
this formula is compatible with the  expression of the inverse metric given in \cite{barros, geiller2}
\bea\label{inverseg}
\text{det}(g) \, g^{ab} \; = \;  (1-\chi^2) \, E^a_i \gamma^{ij}  E^b_j \, ,\quad
\gamma^{ij} \equiv  \delta_{ij} - \frac{\chi_i \chi_j}{1-\chi^2}  \, ,
\eea
due to the properties
\bea
E^a_i = \text{det}(e) e^a_i \, , \quad
\text{det}(g)= (1-\chi^2)\text{det}(e)^2 \, ,\quad
\gamma^{ij} \gamma_{jk}=  \delta^i_k \,  .
\eea
Thus, the identity \eqref{gab}  implies immediately that the metric induced on $\Sigma$ has the same signature as
$\gamma_{ij}$. This latter metric can be easily diagonalized  and its eigenvalues/eigenvectors are easily obtained from
\bea\label{eigen}
 \gamma_{ij} u^j = u_i  \,  \, \text{when} \, \, u \cdot \chi = 0 \, ,\quad
\text{and} \quad
\gamma_{ij} \chi^j = (1-\chi^2) \chi_i \, .
\eea
Therefore, the signature of the metric depends on the sign of $(\chi^2-1)$:  $\Sigma$ is spacelike when $\chi^2<1$ whereas
it inherits a Lorentzian metric when $\chi^2>1$. This clearly explains the presence of $\sigma$ in the constraints algebra \eqref{sigma}
and the nature of the gauge symmetry.
When the symmetry algebra is $\mathfrak{su}(2)$, the space-time is foliated
as usual  into hypersurfaces orthogonal to a timelike vector whereas it is foliated in a space-like direction when the symmetry algebra
is $\mathfrak{su}(1,1)$. This latest case is not conventional but it is the one we are interested in.

	\subsection{Phase space parametrization}
From now on, we will mainly focus on the case $\chi^2 >1$ which has never been studied so far
(we will shortly discuss the case $\chi^2<1$ at the end of this Section).
As the theory admits $\mathfrak{su}(1,1)$ as a gauge symmetry algebra, it
is natural to look for a parametrization of the phase space	adapted to this symmetry. More precisely, we look for conjugate variables
which transform in a covariant way under the Poisson action of the $\mathfrak{su}(1,1)$ generators. In a first part, we exhibit an unique
 $\mathfrak{su}(1,1)$-valued connection which is commutative in the sense of the Poisson bracket. This connection is the $\mathfrak{su}(1,1)$
 analogous of the generalized Ashtekar-Barbero connection defined for $\chi \neq 0$ in \cite{geiller1,geiller2} for instance.
 In a second part, we show that
 it is canonically conjugate to an electric field which transforms as a vector under the action of the first class constraints. Hence, the
 $\mathfrak{su}(1,1)$-connection together with its conjugate electric field provide us with a very useful and natural
 parametrization of the phase space. We finish with computing the action of the vectorial constraints on these variables which transform
 as expected under the action of the generators of diffeomorphisms.

	\subsubsection{The connection}
Now, we address the problem  of finding an $\mathfrak{su}(1,1)$ connection defined by
\bea
{\cal A} = {\cal A}^0 \, J_0 +{\cal A}^1 \, J_1 +{\cal A}^2 \, J_2   \quad \text{with} \quad [J_\alpha,J_\beta]=\epsilon_{\alpha\beta}{}^\tau J_\tau
\eea
which satisfies the following requirements. First, is constructed from the components of $A$ (such that it is commutative in the sense of the Poisson bracket) and the non-dynamical vectors ($\chi$, $u$ and $v$) only. Second it transforms as
\bea\label{transcalA}
\delta_\varepsilon  {\cal A} \; =  \; d\varepsilon + [{\cal A},\varepsilon] \, ,
\eea
under the action of the gauge transformations where $\varepsilon=\varepsilon^\alpha(x)J_\alpha$ is an arbitrary
$\mathfrak{su}(1,1)$-valued function on $\Sigma$. For this relation to make sense, we have to precise the
definition of $\delta_\varepsilon$ in terms of the gauge generators. In particular, we have to establish the link between
the parameter $\varrho \in \mathbb R^3$ entering in the smeared constraint $\tilde{\cal J}(\varrho)$ and the parameter
$\varepsilon$ defining the $\mathfrak{su}(1,1)$ infinitesimal gauge transformations of $\cal A$. From \eqref{projectJ}, it is natural
to expect that
\bea\label{eqforcalA}
\delta_\varepsilon {\cal A} \; = \; \{ \tilde{\cal J}(\varrho) , {\cal A} \} \quad \text{with} \quad
\varepsilon^0= \varrho \cdot n  , \,\, \varepsilon^1= c_1 \varrho \cdot u  , \,\, \varepsilon^2= c_2 \varrho \cdot v \, ,
\eea
where $c_1$ and $c_2$ are functions of $\chi$.  Now, the problem consists in finding the components of $\cal A$ and the
functions $c_1$ and $c_2$ such that $\cal A$ transforms as an $\mathfrak{su}(1,1)$ connection under the action of the
first class constraints.

We are going to propose an ansatz for ${\cal A}$.
As the expressions of the gauge generators are simpler with $\tilde{A}$ instead of $A$ itself,
 we also look for an $\mathfrak{su}(1,1)$ connection $\cal A$ written in terms of $\tilde{A}$.
 This is possible because, when $\chi^2 \neq 1$, $\tilde{A}$ can be uniquely expressed in terms of $A$ and $\chi$ inverting
the relation \eqref{tildeA} as follows:
\bea
A_a \; = \; \tilde{A}_a + \partial_a \chi + \chi \cdot ( \tilde{A}_a + \partial_a \chi) \frac{\chi}{1-\chi^2} \, .
\eea
Inspiring ourselves from the decomposition \eqref{projectJ} of the first class constraints into $\mathfrak{su}(1,1)$ gauge generators,
we propose the following form for the components of $\cal A$:
\bea\label{calA}
{\cal A}^0 = p_0\, (\tilde{A} \cdot n) + q_0 \, , \quad
{\cal A}^1 = p_1 \,(\tilde{A} \cdot u) + q_1 \, , \quad
{\cal A}^2 = p_2\, (\tilde{A} \cdot v) + q_2 \, ,
\eea
where ($p_0,p_1,p_2$) are functions of $\chi $ whereas ($q_0,q_1,q_2$) are one-forms constructed from $d\chi$, $du$ and $dv$ only.

Hence, the problem reduces now in finding the functions ($c_1,c_2$) and ($p_0,p_1,p_2$) together with the one-forms ($q_0,q_1,q_2$)
which solve the equations \eqref{eqforcalA}. These equations can be more explicitly written as
\bea
p_0 \{ \tilde{\cal J}(\varrho), (\tilde{A} \cdot n) \} & = & d(\varrho \cdot n) + c_1( p_2\, (\tilde{A} \cdot v) + q_2)  \varrho \cdot u -
c_2(p_1 \,(\tilde{A} \cdot u) + q_1)  \varrho \cdot v \, ,\\
p_1 \{ \tilde{\cal J}(\varrho), (\tilde{A} \cdot u) \} & = & d(c_1 \varrho \cdot u) + ( p_2\, (\tilde{A} \cdot v) + q_2) \varrho \cdot n -
c_2(p_0\, (\tilde{A} \cdot n) + q_0) \varrho \cdot v \, ,\\
p_2 \{ \tilde{\cal J}(\varrho), (\tilde{A} \cdot v) \} & = & d(c_2 \varrho \cdot v) -
(p_1 \,(\tilde{A} \cdot u) + q_1) \varrho \cdot n+ c_1(p_0\, (\tilde{A} \cdot n) + q_0)  \varrho \cdot u  \, ,
\eea
where each Poisson brackets on the l.h.s. are easily deduced from
\bea
	\{ \tilde{\cal{J}}(\varrho), \tilde{A} \} =  -\frac{1}{\gamma} (1-\chi^2) d \varrho  +
	 \tilde{A} \times \varrho -\frac{1}{\gamma} \chi \times \left( d \chi \times \varrho \right)  +
	 (\tilde{A} \cdot \chi \times \varrho) \chi\, .
\eea
A straightforward calculations show that the previous system reduces to the following three sets of equations:
\begin{eqnarray*}
\begin{array}{ll}
p_0(1-\chi^2) = c_1 p_2 = c_2 p_1 = -\gamma \, , &
 dn + c_1 q_2 u - c_2 q_1 v = 0 \, , \\
p_1 = - p_2 = - c_2 p_0 = {\gamma c_1}/({\chi^2-1})\, , &
d(c_1 u) + q_2 n - c_2 q_0 v +  p_1[(u\cdot d\chi) \chi - (\chi \cdot d\chi) u]/\gamma=0 \, ,\\
p_1=-p_2= c_1 p_0 =-\gamma c_2 /(\chi^2-1)\, , &
d(c_2 v) -q_1 n+ c_1 q_0 u + p_2 [(v \cdot d\chi) \chi - (\chi \cdot d\chi)v]/{\gamma} =0\, .
\end{array}
\end{eqnarray*}
This is clearly an overcomplete set of conditions for the unkowns of the problem. However, an immediate analysis shows that
(up to a simple sign ambiguity), the system admits an unique solution given by
\bea
&&p_0= \frac{\gamma}{\chi^2 -1} \, , \quad p_1= \frac{\gamma}{\sqrt{\chi^2-1}} \, , \quad p_2= -\frac{\gamma}{\sqrt{\chi^2-1}} \, , \\
&&q_0= dv \cdot u \, , \quad q_1 = -\frac{1}{\sqrt{\chi^2-1}} v \cdot dn \, , \quad q_2 = -\frac{1}{\sqrt{\chi^2-1}} u \cdot dn \, ,
\eea
with $c_1=-c_2=\sqrt{\chi^2-1}$.

As a conclusion, let us summarize the main results of this part.
The theory admits  an $\mathfrak{su}(1,1)$ gauge connection
${\cal A}={\cal A}^0 J_0 + {\cal A}^1 J_1 + {\cal A}^2 J_2$ whose components are
\bea
{\cal A}^0 &=& \frac{\gamma}{\chi^2-1} \tilde{A}\cdot n + u \cdot dv \, ,\\
{\cal A}^1 &=& \frac{1}{\sqrt{\chi^2-1}} \left( \gamma \tilde{A}\cdot u -v \cdot dn \right) \, ,\\
{\cal A}^2 &=& -\frac{1}{\sqrt{\chi^2-1}} \left( \gamma \tilde{A}\cdot v + u \cdot dn \right)  \, .
\eea
We have just proved that it transforms as follows
\bea
\delta_\varepsilon {\cal A} = \{\tilde{\cal J}(\varrho), {\cal A}\} = d \varepsilon + [{\cal A},\varepsilon] \quad \text{with} \quad
\varrho= \varepsilon^0 n + \frac{\varepsilon^1 u - \varepsilon^2 v}{\sqrt{\chi^2-1}}
\eea
under the action of the first class constraints. Note that this transformation law is totally consistent with the fact that
\bea
\tilde{\cal J}(\varrho) = {\cal J}_0(\varepsilon^0) + {\cal J}_1(\varepsilon^1) + {\cal J}_2(\varepsilon^2) \, ,
\eea
where the components of $ \tilde{\cal J}$ are the smeared $\mathfrak{su}(1,1)$ generators introduced in \eqref{projectJ}.

\medskip

\noindent
Let us close this analysis with two remarks.

\noindent
First, one can reproduce exactly the same analysis  when $\chi^2 <1$. In that case, one obtains an $\mathfrak{su}(2)$ connection
whose expression is very similar to the previous one obtained for $\mathfrak{su}(1,1)$: everything happens as if one makes the
replacement
$\sqrt{\chi^2-1} \mapsto -\sqrt{1-\chi^2}$ in the components of the connection. The $\mathfrak{su}(2)$-valued connection
is certainly related to the generalized Ashtekar-Barbero connection obtained in  different ways \cite{alexandrov4, les italiens, geiller2}.
In the limit $\chi \rightarrow 0$ with $n$ constant, one recovers the usual Ashtekar-Barbero connection in the time-gauge written in the
orthonormal basis $(n,-u,v)$:
\bea
{\cal A} = {\cal A}^0 \,n - {\cal A}^1 \,u + {\cal A}^2 \, v = \gamma A \, .
\eea

\noindent
Second, by construction, the limit $\chi \rightarrow 0$ does not exist for the $\mathfrak{su}(1,1)$-valued connection.  The analogous of the time
gauge is defined by the limit $\vert \chi \vert \rightarrow \infty$ where the direction $n$ tends to a constant. Let us study this limit, and for simplicity, we assume that the direction $n$ is constant. Starting from the relation
\bea
\tilde{A}_a^i = {}^\gamma \omega_a^{ij} \chi_j +{}^\gamma \omega_a^{0i} - {}^\gamma\omega_a^{0i} \chi_j \chi^i \, ,
\eea
we obtain the following limits for the components of $\cal A$
\bea\label{chiinfinite}
{\cal A}_a^0 \rightarrow - \gamma \; {}^\gamma \omega^{0i}_a n_i \, , \qquad
{\cal A}_a^1 \rightarrow \gamma \; {}^\gamma \omega_a^{ij} n_j u_i \, ,\qquad
{\cal A}_a^2 \rightarrow -\gamma \; {}^\gamma \omega_a^{ij} n_j v_i \, .
\eea
One recognizes the components of the spin-connection in what we could call the ``space-gauge" which would be defined by
the choice $e^a_i n^i = 0$ (instead of $e^a_0=0$ for the usual time gauge).
As a consequence, the limit $\vert \chi \vert \rightarrow \infty$
with $n$ constant is well-defined and consists in a foliation of the space-time ${\cal M}=\Sigma \times \mathbb R$
where the slices $\Sigma$ are orthogonal to the space-like vector $(0,n)$.

	\subsubsection{The electric field }
We follow the same strategy to construct an electric field $\cal E$ which transforms as an $\mathfrak{su}(1,1)$ under the gauge
transformations. More precisely, we are looking for ${\cal E}={\cal E}^0 J_0 +{\cal E}^1 J_1 +{\cal E}^2 J_2 $
which satisfies two conditions. First we require  its components to be constructed from $E$, $\chi$, $u$ and $v$ only and we consider the
natural ansatz
\bea\label{calE}
{\cal E}^0 \; = \; r_0 (E \cdot n) \, , \quad
{\cal E}^1 \; = \; r_1 (E \cdot u) \, , \quad
{\cal E}^2 \; = \; r_2 (E \cdot v) \, ,
\eea
where $(r_0,r_1,r_2)$ are functions of $\chi$ only. Second we require $\cal E$ to transform as a vector
\bea\label{transE}
\delta_\varepsilon {\cal E} \equiv  \{\tilde{\cal J}(\varrho), {\cal E}\} = [{\cal E},\varepsilon] \quad \text{with} \quad
\varrho= \varepsilon^0 n + \frac{\varepsilon^1 u - \varepsilon^2 v}{\sqrt{\chi^2-1}} \, ,
\eea
in adequacy with what has been done in the previous part for the connection.
A simple calculation shows that these conditions implies necessarily
\bea
r_1=\sqrt{\chi^2-1} \, r_0 \, , \quad
r_2=-\sqrt{\chi^2-1} \, r_0 \, ,
\eea
where, at this point, $r_0$ is free because equations \eqref{transE} form a linear system for the unknowns $(r_0,r_1,r_2)$.

\medskip

\noindent
Let us close this analysis with three remarks.

\noindent
First, the free parameter $r_0$ can be fixed requiring in addition that ${\cal E}$ is canonically conjugate to $\cal A$
according to
\bea
\{ {\cal A}^1, {\cal E}^1\} \, = \,
\{ {\cal A}^2, {\cal E}^2\} \, = \, 1 \quad \text{and} \quad
\{ {\cal A}^0, {\cal E}^0\} \,= \, -1  ,
\eea
which easily leads to $r_0 =1/\gamma$.

\medskip

\noindent
Second, it will be useful to express the (inverse of the) induced metric $q^{ab}$ on $\Sigma$ in terms of the $\mathfrak{su}(1,1)$-covariant
electric field. A direct calculation shows that
\bea\label{metricE}
\text{det}(g) \, g^{ab} \, = \; - \gamma^2\,  {\cal E}^{\alpha a} \,   \eta_{\alpha \beta}  \, {\cal E}^{\beta b}   \,  .
\eea
Note that this formula makes very clear that the metric $g_{ab}$ is Lorentzian and its signature is $(-1,+1,+1)$ as we have already
seen in a previous analysis \eqref{eigen}.

\medskip

\noindent
The final remark concerns the $\mathfrak{su}(1,1)$ gauge generators ${\cal J}_\alpha$. It is immediate to see that one can express them
in terms of $\cal A$ and $\cal E$ only as follows
\bea
{\cal J}_\alpha(x) \, J^\alpha \; = \; \partial_a {\cal E}^a(x) \, + \, [ {\cal A}_a(x) \, , \,  {\cal E}^a(x) ] \, .
\eea
We recover the usual Gauss-like form of the constraints, and this expression makes very clear that ${\cal A}$ and ${\cal E}$
transforms respectively as a connection and a vector under the action of the gauge generators.

	\subsubsection{Transformations under diffeomophisms}
As  for the Ashtekar-Barbero connection (or its generalization), we do not expect $\cal A$ to be a fully space-time
connection on $\cal M$. However, it must transform correctly under diffeomorphisms induced on the hypersurface $\Sigma$. To see this is indeed
the case, we first need to identify the generators of  diffeomorphisms  on $\Sigma$. A direct calculation shows that they are given by
the following linear combination of the $\mathfrak{su}(1,1)$ gauge generators and the vectorial constraints:
\bea
{\tilde{\cal{H}}}(N^a) &\equiv &{\cal{H}}(N^a)-\frac{\gamma}{(1+\gamma^2)\chi^2}
\tilde{\cal{J}}(N^a \Omega_a) \, \quad
\text{with} \quad \Omega_a \equiv \gamma {\chi \times A_a} -( A_a \cdot \chi) \chi  \, ,
\eea
which, after some calculations, reduces to
\bea
	{\tilde{\cal{H}}}(N^a) & = & \int d^3 x \, N^a \left( E^b \cdot (\partial_{a} A_{b}- \partial_{b} A_{a})-A_a \cdot \partial_b E^b +\zeta_0 \cdot \partial_a \chi \right) \nonumber \\
	&=& \int d^3 x \, N^a \eta_{\alpha\beta} \, \left( {\cal{E}}^{\alpha b} \cdot (\partial_{a} {\cal{A}}^{\alpha}_{b} - \partial_{b} {\cal{A}}^{\alpha}_{a})
	-{\cal{A}}^{\alpha}_{a} \cdot \partial_b  {\cal{E}}^{\alpha b} \right) \, .
	\eea
Hence, it is immediate to see from this last expression that the constraints ${\tilde{\cal{H}}}(N^a)$ form the algebra of diffeomorphisms.
Furthermore, their actions on ${\cal{A}}$ and $\cal E$ is exactly the lie derivative along the vector field $N^a$:
	\bea
	\{ {\tilde{\cal{H}}}(N^a), {\cal{A}}_b \}=-{\cal{L}}_{N^a} {\cal{A}}_b \, , \quad
	\{ {\tilde{\cal{H}}}(N^a), {\cal{E}}_b \}=-{\cal{L}}_{N^a} {\cal{E}}_b \, .
	\eea
Thus, as announced above, ${\cal{A}}$ is an $\mathfrak{su}(1,1)$-valued connection on $\Sigma$.

	\section{On  the quantization}
We have now all the ingredients to start the quantization of gravity formulated in terms of the $\mathfrak{su}(1,1)$ gauge connection.
Following the standard construction of Loop Quantum Gravity, we assume that quantum states are polymer states, and then
we build the kinematical Hilbert space from holonomies of the connection along edges on $\Sigma$.

\subsection{Quantum states on a fixed graph}
As usual, to any  closed graph $\Gamma \subset \Sigma$ with $N$ nodes and $E$ edges, one associates a kinematical Hilbert
space ${\cal H}_{kin}(\Gamma)$ which is isomorphic to
\bea
{\cal H}_{kin}(\Gamma) \; \simeq \; \left( \text{Fun}[SU(1,1)^{\otimes E}]/SU(1,1)^{\otimes N} ; \, d\mu^{\otimes E} \right) \, ,
\eea
where $d\mu$ is the Haar measure on $SU(1,1)$.
Due to the non-compactness of the gauge group, such a
Hilbert space needs  a regularization to be well-defined (which consists basically in ``dividing'' by the infinite volume of the group).
The details of the regularization of non-compact spin-networks
has been well studied  in \cite{freidel livine}. However, it is well-known that the ``projective
sum" $\oplus_\Gamma {\cal H}_{kin}(\Gamma)$ on the space of all graphs on $\Sigma$ is ill-defined and, up to our knowledge,
no one knows how to construct a non-compact Ashtekar-Lewandowski measure.
Thus, only the kinematical Hilbert space on a fixed graph $\Gamma$ is mathematically well-defined and we limit the study of quantum
states as elements of ${\cal H}_{kin}(\Gamma)$ only.
Hence, a quantum state is a function $\psi_\Gamma [A] \equiv f(U_1,\cdots,U_E)$ of the holonomies
\bea
U_e \; \equiv \; P\exp \int_e {\cal A} \; \in SU(1,1)
\eea
along the edges $e$ of $\Gamma$. The electric field ${\cal E}$ is promoted as an operator whose action on $\psi_\Gamma$ is formally
given by
\bea
\hat{\cal E}{}^a_i(x) \, \psi_\Gamma[A] \; = \; i \ell_p^2\frac{\delta}{\delta {\cal A}_a^i(x)}  \psi_\Gamma[A]\, ,
\eea
where $\ell_p$ is the Planck length.
Note that the flux of $\cal E$ across a surface is a well-defined operator on  ${\cal H}_{kin}(\Gamma)$: it acts as a vector field on
the space of $SU(1,1)$ functions.

The Peter-Weyl theorem implies that $\psi_\Gamma$ can be formally decomposed as follows
\bea
\Psi_\Gamma[A] \; = \; \sum_{s_1,\cdots,s_E} \text{tr}\left( \tilde{f}(s_e) \, \bigotimes_{e=1}^E  \pi_{s_e}(U_e) \right)
\eea
where
\bea
\pi_s: SU(1,1) \rightarrow \text{End}(V_s) \,  \quad \text{and} \quad
\tilde{f} \in \bigotimes_{e=1}^N V_{s_e}^* \, .
\eea
The sum runs over  unitary irreducible representations of $SU(1,1)$ labelled generically by $s_e$. We used the notation
 $V_{s_e}$ for the modulus of the representation, $V_{s_e}^*$ for its dual, and $\text{tr}$ denotes the pairing between
$\otimes_{e} V_{s_e}$ and its dual $\otimes_{e} V_{s_e}^*$.
 Due to the gauge invariance of $\psi_\Gamma$,
 the Fourier modes $\tilde{f}$ are in fact $SU(1,1)$ intertwiners and the expression of $\psi[A]$ needs a regularization
 to be well-defined \cite{freidel livine}. Furthermore, unitary irreducible
 representations of $SU(1,1)$, which are classified into the two discrete series (both labelled with integers) and the continuous series
 (labelled with real numbers), are infinite dimensional (see \cite{Rhul} for a review on representations theory of $\mathfrak{su}(1,1)$).

 \subsection{Area operators}

 Thus, edges of $SU(1,1)$ spin-networks can be colored with discrete or real numbers. The geometrical interpretation is clear: these two different types of colors label edges which are normal to either time-like or space-like surfaces.  To see how to link the representations to
 the time-like or space-like natures of the surfaces, we have to compute the spectrum of the area operators in terms of the quadratic
 Casimir of $\mathfrak{su}(1,1)$. For that purpose, we start with the expression \eqref{metricE} of the inverse metric $g^{ab}$ that we contract twice with the normal $n_a$ to a given surface $S$. This leads to  the formula
 \bea
 \text{det}(g) \, n^2 \, = \, -\gamma^2 \, (n_a{\cal E}^{a\alpha} ) \, \eta_{\alpha\beta} \,  ({\cal E}^{b\beta} n_b) \, ,
 \eea
where $n^2=n_an_b g^{ab}$. Hence, the determinant of the induced metric $h$ on the surface $S$ is given by
\bea
 \text{det}(h) \, = \, -\gamma^2 (n_a{\cal E}^{a\alpha} ) \, \eta_{\alpha\beta} \,  ({\cal E}^{b\beta} n_b)  \, .
\eea
As a consequence, the action of the area operator $\hat{S}$, punctured by an edge $e$ of the graph $\Gamma$ colored by
a representation $s_e$, on ${\cal H}_{kin}(\Sigma)$ is diagonal and its eigenvalue $S(s)$ is given by the equation
\bea
S(e)^2 \; = \;  - i^2 \gamma^2 \ell_p^4 \;  \pi_e(J_1^2 + J_2^2 -J_0^2) \; = \; \gamma^2 \ell_p^4 \, \pi_e (C)
\eea
where $\pi_e(C)$ is identified with  the unique eigenvalue of the Casimir tensor $C\equiv -J_0^2 + J_1^2 + J_2^2$ in the representation $s_e$.
Obviously, the evaluation $\pi_e(C)$ depends on the nature discrete $(s_e=j_e \in \mathbb N)$ or continuous $(s_e \in \mathbb R)$
of the representation according to
\bea
\pi_{j_e}(C)= j_e(j_e+1) \quad \text{and} \quad \pi_{s_e}(C)=-(s_e^2 + \frac{1}{4}) \,.
\eea
We deduce immediately that $S(e)^2$ is positive when $e$ is colored with a discrete representation whereas $S(e)^2$ is negative when $e$
is colored with a representation in the continuous series. As a consequence,
the area operator of any space-like surface has a discrete spectrum and
the area operator of  any time-like surface has a continuous spectrum. Furthermore, the spectrum of space-like areas is in total agreement of the usual spectrum in Loop Quantum Gravity.  Note that a very similar result has been recently derived in the context of twisted
geometries \cite{rennert}.

	\section{Discussion}
In this paper, we have  formulated gravity as an $SU(1,1)$ gauge theory. We have started with the Hamiltonian
formulation of the fully Lorentz invariant Holst action on a space-time manifold of the form ${\cal M}=\Sigma \times \mathbb R$.
Then we have considered a partial gauge fixing which reduces the internal
$\mathfrak{sl}(2,\mathbb C)$ gauge symmetry to $\mathfrak{su}(1,1)$. The 3-dimensional slice $\Sigma$ inherits a
Lorentzian metric of signature $(-,+,+)$.
The partial gauge fixing relies on the introduction on an external
non-dynamical vector field $\chi$ which measures the normal of the hypersurface $\Sigma$ but it plays in fact no physical role at the end of the process.

Next we found that the phase space
of the partially gauge fixed theory is well-parametrized by a pair $({\cal A},{\cal E})$ formed with an $\mathfrak{su}(1,1)$-valued connection on
$\Sigma$ and its canonically conjugate electric field whose components can be identified to vectors in the flat (2+1) Minkowski space-time.
The phase space comes with first class constraints: the Gauss constraints which generate $\mathfrak{su}(1,1)$ gauge transformations,
the vectorial constraints which have been shown to generate diffeomorphisms on $\Sigma$ and the usual scalar constraint that we have
not studied in this paper.

Finally, we have explored the quantization of the theory studying some aspects of the kinematical Hilbert space
${\cal H}_{kin}(\Gamma)$ on a fixed given graph $\Gamma$ which lies on $\Sigma$. Due to the non-compactness of the gauge group $SU(1,1)$,  ${\cal H}_{kin}(\Gamma)$ needs a regularization to be well-defined and the projective sum over all possible graphs is not
under control. This is why we restrict our study to the quantization on a fixed graph only. We compute the spectrum of the area operators
acting on  ${\cal H}_{kin}(\Gamma)$  and  found that the spectrum is discrete for space-like surfaces and continuous for time-like surfaces. Furthermore, the usual quantization of the Holst action in the time-gauge ($\chi=0$) and the new quantization presented here and based to another totally inequivalent partial gauge fixing $(\chi^2 >1)$ lead to exactly the same spectrum of the area operator (on space-like surfaces) at the kinematical level.
This strongly suggests that the time gauge introduces no anomaly in the quantization of gravity, at least at the kinematical level, as it
was already underlined in \cite{geiller2} in a  different situation.

\medskip

This  formulation of gravity seems very interesting because it offers another point of view on the quantization  of gravity
in four dimensions. Now, we have a description of the kinematical quantum states of gravity not only on space-like surfaces $\Sigma$
but also on time-like surfaces (only remains the description of the quantum states on null-surfaces, what we hope to study in the
future). Hence, with those space-like and time-like kinematical quantum states, we are not far from having a fully covariant description of
quantum gravity. In that respect, it would be very instructive to make a contact between these two canonical quantizations and
spin-foam models for covariant quantum gravity. Furthermore, if we understand how to ``connect" the time-like and the space-like
kinematical quantum states, we could open a new and promising way towards a better understanding of the dynamics in Loop Quantum
Gravity.

\begin{figure}
	\centering
	\subfigure[Time-like slices] { \label{fig:a}    \includegraphics[width=0.45\columnwidth]{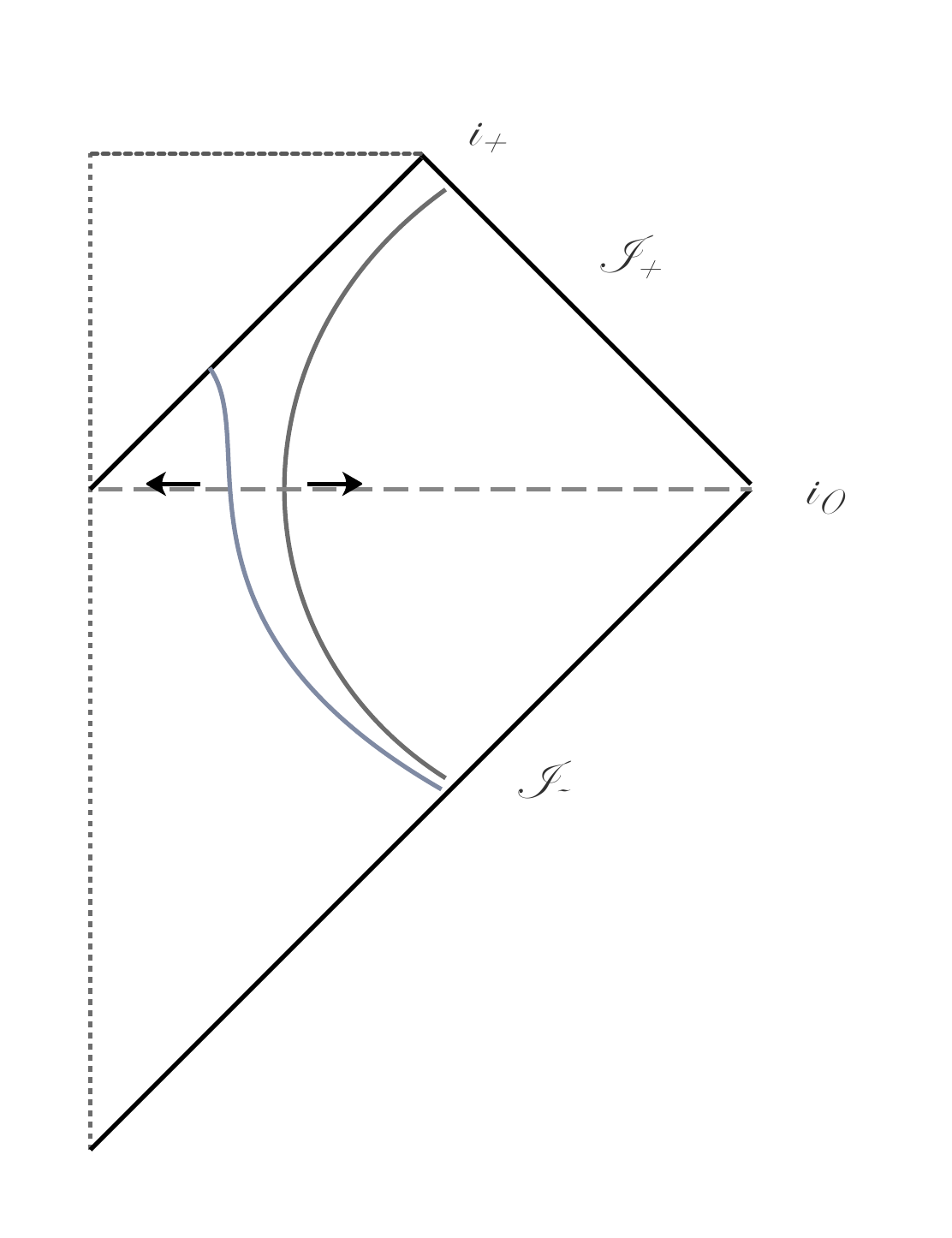} }
	\subfigure[Space-like slices] { \label{fig:b}    \includegraphics[width=0.45\columnwidth]{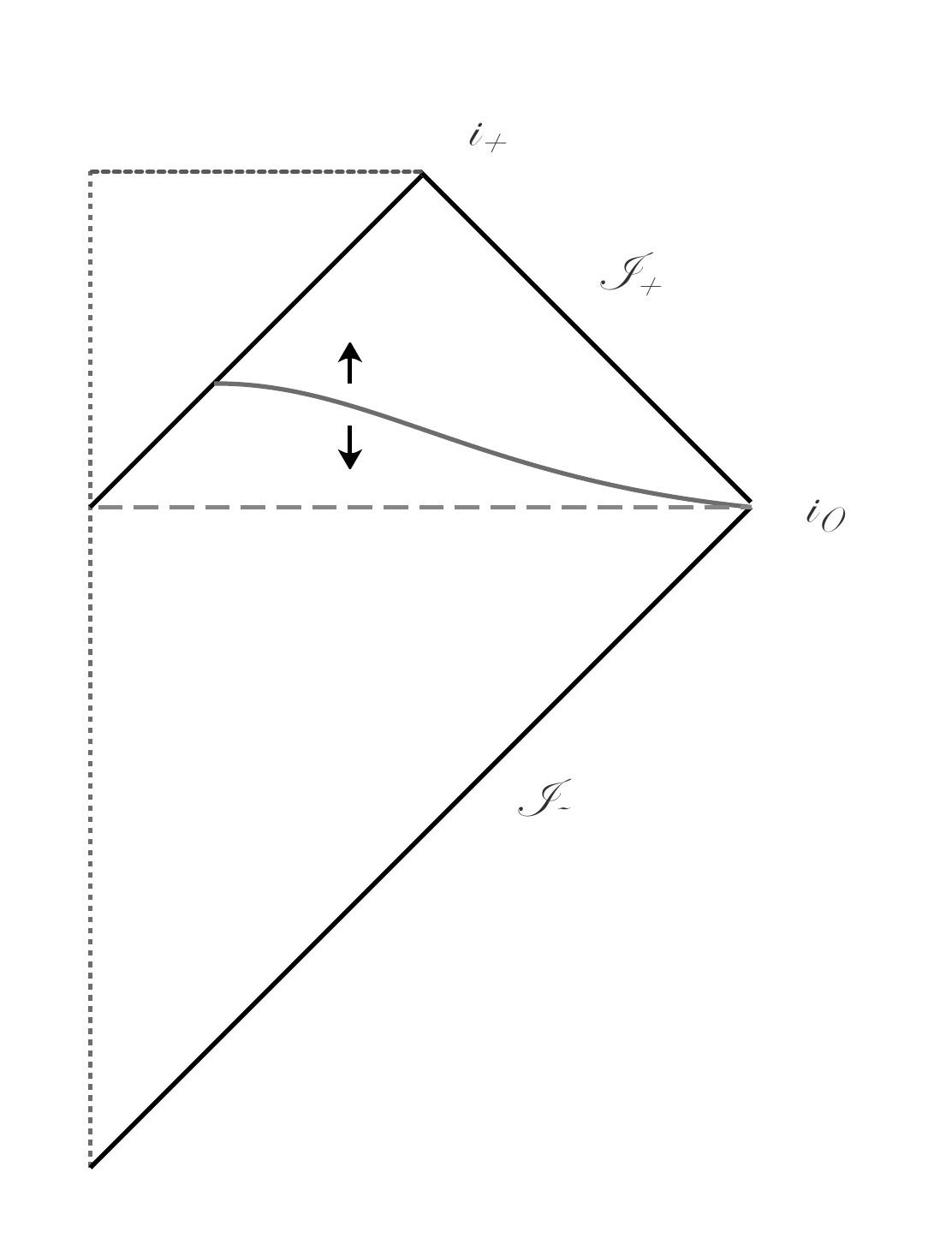} }
	\caption{Different Hamiltonian slicings of a spherical black hole space-time. The picture (b) represents the usual slicing in terms of
	space-like hypersurfaces which leads to the effective $SU(2)$ Chern-Simons description of the black hole: In that case,  the horizon
	appears as a boundary of $\Sigma$. In the picture (a), we have represented two slicings of the black hole space-time where $\Sigma$
	are Lorentzian hypersurfaces: these gauge choices would lead to new descriptions of black holes in Loop Quantum Gravity.
	In particular, the slicing which does not cross the horizon is interesting in view of a holographic description of black holes in the
	frame of Loop Quantum Gravity.}
	\label{fig}
\end{figure}

It is also interesting to notice that the Hamiltonian constraint in the formalism where $\Sigma$ is space-like becomes a component of the vectorial constraints in the formalism where $\Sigma$ is time-like. The reverse is also true. As we know very well how to quantize the
vectorial constraints on the kinematical Hilbert space, we  think again that understanding the relation between these two Hamiltonian
quantizations could lead us to a solution of the Hamiltonian constraint. We hope to study these questions related to the quantum
dynamics in the future.

Beside, we deeply think that this new formulation will allow us to understand better the physics of quantum black holes in Loop Quantum Gravity. In the usual treatment \cite{BH1,BH2,BH3,BH4,BH5,BH6,BH7,BH8}, black holes are considered as isolated horizons and they appear as boundary of a 3 dimensional
space-like hypersurface $\Sigma$. Their effective dynamics has been shown to be governed by an $SU(2)$
Chern-Simons theory whose quantization leads to the construction and the counting of the quantum microstates for the black holes.
With the $\mathfrak{su}(1,1)$ formulation of gravity, it is now possible to start a Hamiltonian quantization of gravity  where $\Sigma$ is time-like. Naturally, one would expect that quantizing black holes with space-like or time-like slices would lead to two equivalent
descriptions of the black hole microstates. At first sight, we would say that, starting with a time-like slicing, one would get an $SU(1,1)$
Chern-Simons theory as an effective dynamics for the spherical black hole for instance. Thus, we can ask the question how an $SU(1,1)$
and an $SU(2)$ Chern-Simons theories could provide two equivalent Hilbert spaces when they are quantized. This may be possible when
$\gamma$ becomes complex and equal to $\pm i$ because, in that case, we expect the two gauge group of the Chern-Simons
theories to become the same Lorentz group. This would give one more argument in favor of the analytic continuation procedure
introduced and studied in \cite{analytic0,analytic1,analytic2,analytic3,analytic4}.
However, this idea might be too naive because, on a time-like slicing, the black hole does not appear as a boundary anymore and a particle
leaving on the slice $\Sigma$ now cross the horizon and does not see any border.
To finish, this new formulation of Loop Quantum Gravity opens the possibility
to define a kind of ``holographic" description for black holes in the framework of Loop Quantum Gravity as shown in the picture Fig. \ref{fig}
above. We hope to
study all these very intriguing aspects related to black holes in a future work \cite{holography}.

\acknowledgements
We are particularly indebted to Alejandro Perez and Simone Speziale for numerous discussions on this project. K.N. want also to thank
Jibril Ben Achour who initially collaborated on this project.

	\appendix
	\section{``Time'' vs. ``Space" gauge in the Holst action}
	The very well-known ``time" gauge refers to the condition $e^0_a$ which breaks $\mathfrak{sl}(2,\mathbb C)$ into $\mathfrak{su}(2)$
	in the Holst action.
It corresponds to taking a slicing $\Sigma \times \mathbb R$ of the space-time where the hypersurfaces $\Sigma$ are space-like.
In fact, one can easily generalize   the time gauge by considering instead
the condition $e_a^\mu n_\mu=0$ where $n_\mu$ is a given fixed vector. When $n_\mu$ is time-like, the slices $\Sigma$ are space-like
(as for the time gauge where $n_\mu=\delta_\mu^0$) whereas the slices are time-like when $n_\mu$ is space-like. We want to study
thus latter case in this appendix. To simplify the analysis, we assume (without loss of generality) that $n_\mu=\delta_\mu^3$.

We are going to show that the Hamiltonian analysis of the Holst action such a  gauge leads to a phase space which corresponds to the limit $\vert \chi \vert \rightarrow \infty $ and $n_i \rightarrow \delta_i^3$. First, we notice that the only non vanishing components of
$\pi^a_{IJ}$ are $E^a_\alpha \equiv \pi^a_{\alpha 3} $ with $\alpha \in (0,1,2)$. It is immediate to check that the simplicity constraints ${\cal{C}}^{ab} \approx 0$ are satisfied. In this gauge, it is ``natural" to choose the third direction to be the ``time" parameter because of
the slicing. Hence,  the ``symplectic" term (in the third direction) in the Holst action  involves only the  component $\omega_a^{\alpha 3}$ of
the spin-connection  (with $\alpha \in (0,1,2)$ and $a \in (0,1,2)$ also) according to the formula
	\bea
	\gam {\pi}^a_{IJ} \, \partial_3 {\omega}_a^{IJ}=E^a_\alpha \partial_3{A}_a^\alpha \, , \qquad \text{where} \quad
			A_a^\alpha \equiv {}^{\gamma} \!\omega_a^{\alpha 3} \, .
	\eea
Hence, the connection $A$ is clearly the  variable canonically conjugate  to $E$.
Finally, one  shows that the resolution of the second class constrains ${\cal{D}}^{ab} \approx 0$ leads to the following expression
for the gauge generators
\bea
	{\cal{J}}_{0}& =& -\frac{1}{\gamma} \partial_a E^{a0}-A_a^{1}E^{a2}+A_a^2E^{a1} \, ,\\
	{\cal{J}}_{1}& = &-\frac{1}{\gamma} \partial_a E^{a2}+A^0E^1-A^1E^0 \, ,\\
  {\cal{J}}_{2}& =& \frac{1}{\gamma} \partial_a E^{a}1-A^0E^2+A^2E^0 \, .
	\eea
They satisfy the constraints algebra
	\bea
	\{ {\cal{J}}_0, {\cal{J}}_1 \}\,=\, {\cal{J}}_2 \, , \qquad
	\{ {\cal{J}}_0, {\cal{J}}_2 \}\,=\, -{\cal{J}}_1 \, , \qquad
	\{ {\cal{J}}_1, {\cal{J}}_2 \} \,=\, -{\cal{J}}_0 \, ,
	\eea
which is nothing by the $\mathfrak{su}(1,1)$ algebra. At this point, it is not difficult to see that the associated covariant connection
has the following components
	\bea
	{\cal A}_a^0 = - \gamma \; {}^\gamma \omega^{03}_a \, , \qquad
	{\cal A}_a^1 = \gamma \; {}^\gamma \omega_a^{23} \, ,\qquad
	{\cal A}_a^2 = -\gamma \; {}^\gamma \omega_a^{13} \, .
	\eea
We recover as announced the same expression of the $\mathfrak{su}(1,1)$-valued connection in the limit $\vert \chi \vert \rightarrow \infty$
 \eqref{chiinfinite} a part that we have interchanged the components $0$ and $3$ of space-time indices.

\end{document}